\documentclass[prc,amsmath,twocolumn,showkeys,showpacs,superscriptaddress]{revtex4}
\pdfoutput=1
\usepackage{gensymb}
\usepackage{graphicx,color}
\usepackage{amssymb}
\usepackage{enumerate}
\usepackage{verbatim}
\usepackage{natbib}
\usepackage{float}
\usepackage[outdir=./]{epstopdf}

\begin{document}
  \newcommand {\nc} {\newcommand}
  \nc {\Sec} [1] {Sec.~\ref{#1}}
  \nc {\IR} [1] {\textcolor{red}{#1}} 
  \nc {\IB} [1] {\textcolor{blue}{#1}} 
  \nc {\IG} [1] {\textcolor{green}{#1}}

\title{Uncertainty quantification due to optical potentials in models for (d,p) reactions }

\author{G.~B.~King}
\affiliation{National Superconducting Cyclotron Laboratory, Michigan State University, East Lansing, MI 48824}
\affiliation{Department of Physics and Astronomy, Michigan State University, East Lansing, MI 48824-1321}
\author{A.~E.~Lovell} 
\affiliation{National Superconducting Cyclotron Laboratory, Michigan State University, East Lansing, MI 48824}
\affiliation{Department of Physics and Astronomy, Michigan State University, East Lansing, MI 48824-1321}
\affiliation{Theoretical Division, Los Alamos National Laboratory, Los Alamos, NM 87545}
\affiliation{Center for Nonlinear Studies, Los Alamos National Laboratory, Los Alamos, NM 87545}
\author{F.~M.~Nunes}
\email{nunes@nscl.msu.edu}
\affiliation{National Superconducting Cyclotron Laboratory, Michigan State University, East Lansing, MI 48824}
\affiliation{Department of Physics and Astronomy, Michigan State University, East Lansing, MI 48824-1321}

\date{\today}


\begin{abstract}
\begin{description}
\item[Background:]  
Recent work has studied the uncertainty in predictions for A(d,p)B reactions using the distorted-wave Born approximation (DWBA), coming from the parameterization of the effective dA interactions [Lovell {\it et al.}, Phys. Rev. C 95, 024611]. There are different levels of sophistication in reaction theories for one-nucleon transfer reactions, including the adiabatic wave approximation (ADWA) which takes deuteron breakup into account to all orders.
\item[Purpose:] 
In this work, we quantify the uncertainties associated with the ADWA method that come from the parameterization of the NA interactions, and compare ADWA with DWBA.
\item[Method:] 
Similarly to Lovell {\it et al.}, Phys. Rev. C 95, 024611, we use nucleon elastic-scattering data on a wide variety of targets, at the appropriate incoming and outgoing energies, to constrain the optical potential input to the ADWA theory.  Pulling from the $\chi^2$-distribution, we obtain 95\% confidence plots for the elastic distributions. From the resulting parameters, we predict 95\% confidence bands for the (d,p) transfer cross sections. Results obtained with the standard uncorrelated $\chi^2$ are compared to those using the correlated $\chi^2$ of Lovell {\it et al.}, Phys. Rev. C 95, 024611. We also repeat the DWBA calculations for the exact same reactions for comparison purposes.
\item[Results:] 
We find that NA elastic scattering data provides a significant constraint to  the interactions, and, when the uncertainties are propagated to the transfer reactions using ADWA, predictions are consistent with the transfer data.  
\item[Conclusions:] 
The angular distributions for ADWA differ from those predicted by DWBA, particularly at small angles. As in Lovell {\it et al.}, Phys. Rev. C 95, 024611, confidence bands obtained using the uncorrelated  $\chi^2$-function are unrealistically narrow and become much wider when the correlated $\chi^2$-function is considered. For most cases, the uncertainty bands obtained in ADWA are narrower than DWBA, when using elastic data of similar quality and range. However, given the large uncertainties predicted from the correlated $\chi^2$-function, at this point, the transfer data cannot discriminate between these two methods. 
\end{description}
\end{abstract}

\keywords{uncertainty quantification, nucleon elastic scattering, transfer nuclear reactions, optical potential fitting}

\maketitle

\section{Introduction}
\label{intro}

Advances in rare isotope facilities around the world are leading to unique beam intensities and detector systems that will allow the collection of a wide array of reaction data  on  nuclei far from stability, with higher precision than ever before. As these experimental advances take place, one needs to ask whether the theory needed for the interpretation of this high quality data is adequate. Much effort has been put into benchmarking standard approximations in the field of direct reaction theory (e.g. \cite{Deltuva2007,Nunes2011,Capel2012,Chazono2017,Yoshida2018}).
In addition,  new approaches are being developed  that allow a  more microscopic foundation for the input needed  (e.g. \cite{Toyokawa2015,Rotureau2017,Kumar2017,burrows2018,gennari2018}). 
However, the important question remains: for a given reaction model, what is the uncertainty in the theoretical prediction for the cross section?  In this work, we address this question in connection to one-nucleon (d,p) reactions.

While there are some {\it ab initio} approaches to reactions, most direct-reaction theories resort to retaining only a few degrees of freedom that are essential to describe the relevant reaction mechanisms. Thus, the input for describing  transfer A(d,p)B reactions are effective interactions, the so-called optical potentials,  for the N-A system and/or the d-A system, depending on whether one uses a true three-body approach or a perturbative approach such as the Born series \cite{ReactionsBook}. Ambiguities in the optical potentials represent the single most important source of uncertainty in reaction theory. Efforts to develop these effective interactions from the underlying NN force are underway (e.g. \cite{Rotureau2017}), yet there are still many problems  that need to be resolved before these microscopically-derived effective interactions can be used in the interpretation of reaction data. Most commonly, optical potentials are generated from fitting elastic (and other) data. Global parameterizations are developed for a range of scattering energies and target masses, based on stable nuclei for which data is abundant (e.g. \cite{bg69,ch89,kd2003}). Despite the unknown uncertainties, especially in extrapolating away from stability, global potentials have been widely used and are considered by most to be the best available option.

In order to move the field forward, it is critical to quantify uncertainties and develop tools that enable us to improve predictions made by reaction theory. 
The standard way to determine uncertainties involves using the covariance matrix and exploring the $\chi^2$-function around the best fit.
The study performed  in \cite{Lovell2017} inspected the $\chi^2$-function around minima and determined confidence bands for  angular distributions  for  elastic, inelastic and (d,p) cross sections at energies in the range of $5-25$ MeV/u, on targets with mass $A=12-208$. A comparison was performed between results obtained with the standard $\chi^2$-function and  a correlated $\chi^2$-function. The results in \cite{Lovell2017} demonstrate that the correlated $\chi^2$ function produces parameterizations that are more physical, yet the confidence bands are significantly broader.  The (d,p) reactions included in the study of \cite{Lovell2017} were performed within the distorted-wave Born approximation (DWBA). One of the goals of this work, is to revisit the study in \cite{Lovell2017} with an upgraded reaction model.

In the area of nuclear reactions, there are many alternative models that have additional physics included. The breakup of the deuteron, thought to play an important role in (d,p) reactions, is  treated in a crude manner in DWBA.  Methods such as CDCC \cite{austern1987}, and its simpler relative ADWA \cite{Johnson1974}, provide non-perturbative approaches based on a three-body Hamiltonian. However, Occam's razor tells us that one should opt for the simplest model that is compatible with the data. Is the (d,p) transfer data able to discriminate between DWBA and ADWA, or even reject one of the models? The answer to this question relies on the ability to quantify the uncertainties arising from the theory itself. In this work, we will quantify the uncertainties due to the optical potentials in (d,p) reactions, within both DWBA and ADWA, using the same fitting philosophy. We will then be able to make a meaningful comparison between DWBA and ADWA predictions and determine the level of sophistication required for describing the (d,p) data.

In Section \ref{theory}, we summarize the reaction theory used in this work as well as the statistical methods used. In Section \ref{results}, we present the results obtained when using ADWA and DWBA, as well as a comparison between the two theories. Conclusions are then drawn in Section \ref{conclusions}.

\section{Theory and implementation summary}
\label{theory}

\subsection{Reaction Theory}

Due to the loosely bound nature of the deuteron, it is common to describe A(d,p)B reaction starting from a three-body Hamiltonian of $n+p+A$:
\begin{eqnarray}
\mathcal{H}_{3B}=T_R+T_r+U_{nA}+U_{pA}+V_{np}\;,
\label{eq-h3b}
\end{eqnarray}
where the pairwise interactions $U_{nA}$ and $U_{pA}$ are  effective interactions describing the main features of the nucleon-target systems,  and $V_{np}$ is the known NN force.  $T_R$ and $T_r$ are the two-body kinetic energy operators for the deuteron-target and n-p  systems.

The exact T-matrix amplitude for  $A(d,p)B$  can be written in the post form as:
\begin{eqnarray}
\label{eq-tmatrix}
T=\langle \phi_{nA}\chi^{(-)}_{pB}|V_{np}+\Delta|\Psi^{(+)}\rangle,
\end{eqnarray}
where $\phi_{nA}$ describes the final neutron bound state, $\chi_{pB}$ is the proton distorted wave, and $\Delta=U_{pA}-U_{pB}$ is the remnant term which is negligible for reactions on intermediate and heavy masses. $\Psi$ corresponds to the exact three-body wavefunction in the incident channel, but Johnson and Tandy realized that this wavefunction would only be needed within the range of $V_{np}$ \cite{Johnson1974}. This led to the choice of using the Weinberg basis to expand the three-body wavefunction. If one then makes the adiabatic approximation, neglecting the excitation energy of the $np$ system in the reaction and retains only the first term in the Weinberg expansion, one arrives at a simple form for the T-matrix:
\begin{eqnarray}
\label{eq-tmatrix-ad}
T=\langle \phi_{nA}\chi^{(-)}_{pB}|V_{np}| \phi_{np} \chi^{ad}_d \rangle,
\end{eqnarray}
where the adiabatic wave $\chi^{ad}_d$ is generated from the effective adiabatic potential:
\begin{eqnarray}
U_{AD}=-\langle \phi_0(\textbf{r})|V_{np}\left(U_{nA}+U_{pA} \right)|\phi_{0}(\textbf{r})\rangle,
\label{eq-pot}
\end{eqnarray}
with $\phi_0$ being the first Weinberg eigenstate.

The method of Johnson and Tandy, referred to as the adiabatic wave approximation (ADWA) has been discussed in detail in \cite{nguyen2010}  and was recently extended to incorporate non-local interactions \cite{Titus2016}. ADWA has been been tested against exact Faddeev calculations \cite{Nunes2011}, and results demonstrate its validity for deuteron energies in the range $E=20-40$ MeV.

As apposed to ADWA, the 1-step distorted-wave Born approximation (DWBA), replaces the full incoming wavefunction by the deuteron elastic scattering channel:
\begin{eqnarray}
\label{eq-tmatrix-dwba}
T=\langle \phi_{nA}\chi^{(-)}_{pB}|V_{np}+\Delta| \phi_{np}\chi_{dA} \rangle,
\end{eqnarray}
where $\phi_{d}$ describes the deuteron bound state and $\chi_{dA}$ is the deuteron distorted wave, obtained with the optical potential $U_{dA}$ typically fit to deuteron elastic scattering. In DWBA, deuteron breakup is only included implicitly through the deuteron elastic channel.

The transfer calculations shown in this study include finite-range effects and neglect the remnant term. For the cases considered, the remnant term contributes by less than 3 \%.
For the deuteron bound state and the operator $V_{np}$ in Eq.\ref{eq-tmatrix-ad} and Eq.\ref{eq-tmatrix-dwba}, we  fix the NN interaction to Reid \cite{reid}. The neutron final bound state is described by a Woods-Saxon interaction with standard radius $r=1.25$ fm and diffuseness $a=0.65$ fm, and with the depth adjusted to reproduce the experimental separation energy in the corresponding (A+1) system.

\subsection{Fitting procedure and uncertainty bands}

The optical model describes the elastic scattering of a projectile-target combination in terms of a  $U_{opt}$ effective potential. Given a set of parameters $\textbf{x}$, the optical model makes predictions for the angular distribution $m(\textbf{x},\theta)$ of elastic scattering. The tradition in our field is to minimize the standard (uncorrelated) $\chi^2$ function, which is the sum of the square of the residuals, {the difference between} the differential cross sections predicted by the model $m(\textbf{x},\theta_i)$ and the data $d_i$  for elastic scattering measured at $M$ angles $\theta_i$, with experimental errors $\sigma_i$: 
\begin{equation}
\chi ^2 _{UC}= \frac{1}{M}\sum \limits _{i=1} ^M \frac{(m(\textbf{x},\theta_i) - d_i)^2}{\sigma _i ^2},
\label{eq:chi2}
\end{equation}
In minimizing this $\chi^2$, one finds the best-fit set of parameters $\hat{\textbf{x}}$. 
A large number of parameter sets can then be pulled from the $\chi^2$ distribution around the minimum and run through the optical model to determine a corresponding set of differential cross sections.  Then, 95\% confidence bands are defined by removing the highest 2.5\% and lowest 2.5\% of the predicted values for those cross sections at each angle.  

Typically, the optical potential used to describe a light projectile impinging on a target is parameterized in terms of \cite{bg69}:  
i) a volume real part of Woods-Saxon form with parameters $V, r, a$ for the depth, radius and diffuseness;
ii) a volume imaginary term of Woods-Saxon form, with parameters $W_v, r_v, a_v$;
iii) a surface imaginary term, proportional to the derivative of a Woods-Saxon form, with parameters $W_s, r_s, a_s$;
iv)  a spin-orbit term and v) when the projectile has charge, a Coulomb force. Note that the parameters associated with the spin-orbit and the Coulomb are kept fixed in our procedure, while all others are in principle allowed to vary to find the best minimum.

The $\chi^2$ function in Eq. \ref{eq:chi2} assumes that the model predictions at angles $\theta_i$ and $\theta_j$ are independent. As argued in \cite{Lovell2017}, due to the angular momentum decomposition, all angles in our model are correlated. We have thus considered the following correlated $\chi^2$ function \cite{Lovell2017}:
\begin{equation}
\chi^2_C =  \frac{1}{M}\sum \limits _{i=1}^M \sum \limits _{j=1}^M W_{ij}(m(\textbf{x},\theta_i)-d_i)(m(\textbf{x},\theta_j)-d_j).
\label{eq:chi2c}
\end{equation}
$W_{ij}$ are the matrix elements of $\mathbb{W} = (\mathbb{C}_m + \Sigma)^{-1}$, where $\mathbb{C}_m$ is the model covariance matrix, which is assumed to describe the correlations between calculated cross section values at different angles, and $\Sigma$ is a diagonal matrix with $\sigma _i ^2$ on the diagonals. (When $\mathbb{C}_m=0$, Eq. \ref{eq:chi2c} reduces to Eq. \ref{eq:chi2}.) Once the best fit associated with $\chi^2_C$ is obtained, one can again construct a 95\% confidence band, pulling a large set of parameters from this new correlated distribution and removing the highest and lowest 2.5\% of the predicted values for the observable of interest. 
\begin{table}[t]
\begin{center}
\begin{tabular}{|c|c|c|}
\hline Reaction &Energy (MeV) &Reference\\ \hline
$^{48}$Ca(p,p) &12 &\cite{48cap12}\\ \hline
$^{48}$Ca(n,n) &12  &\cite{48can12}\\ \hline
$^{48}$Ca(p,p) &25 &\cite{48cap25}\\ \hline
$^{48}$Ca(d,d) &23.2 &\cite{48cad23} \\ \hline
$^{90}$Zr(p,p) &12.7 &\cite{90zrp12}\\ \hline
$^{90}$Zr(n,n) &10 &\cite{90zrn10} \\ \hline
$^{90}$Zr(p,p) &22.5 &\cite{90zrp22} \\ \hline
$^{90}$Zr(p,p) &9.018 &\cite{90zrp9} \\ \hline
$^{90}$Zr(n,n) &24 &\cite{90zrn10} \\ \hline
$^{90}$Zr(d,d) &23.2 &\cite{48cad23} \\ \hline
$^{208}$Pb(p,p) &16 &\cite{208pbp16} \\ \hline
$^{208}$Pb(n,n) &16.9 &\cite{208pbn16} \\ \hline
$^{208}$Pb(p,p) &35 &\cite{208pbp35} \\ \hline
$^{208}$Pb(d,d) &28.8 &\cite{208pbd28} \\ \hline
\end{tabular}
\caption{References to the experimental data used for the elastic-scattering fits, including the energy of the data.}
\label{tab:data}
\end{center}
\end{table}

\subsection{Data and fitting protocol}

An important goal in this work is to obtain 95\% confidence bands for (d,p) angular distributions, reflecting the ambiguities in the optical potentials.
For this purpose, we use elastic-scattering data to constrain all of the optical potentials that are input to the transfer reaction model. We carefully reviewed the literature and compiled those cases for which there is nA, pA and dA elastic scattering data, at the relevant energies (both for the incoming channel and the outgoing channel) as well as the corresponding A(d,p)B data. 
Note that, for constraining the optical potential in the outgoing proton channel, we fit proton elastic scattering on the closed-shell systems ($^{48}$Ca, $^{90}$Zr, $^{208}$Pb), for which there is more data, than for the $(A+1)$  systems ($^{49}$Ca, $^{91}$Zr, $^{209}$Pb). We then rescale the radius appropriately.
Wide angular distributions and low errors bars were important considerations in our selection.  In the end, the cases considered are provided in Table \ref{tab:data}, along with the sources used for the data, through \cite{nndc}. Experimental error bars were taken directly from \cite{nndc}.

We used modified versions of {\sc sfresco} \cite{fresco} to perform  $\chi^2$ minimizations for the angular distributions generated with the optical model, {\sc fresco} \cite{fresco} for elastic-scattering confidence bands, and {\sc nlat} \cite{nlat}  to determine the transfer cross section in DWBA and ADWA respectively. Confidence bands were produced with 800 pulls from the $\chi^2$ distributions.

\section{Results}
\label{results}

\subsection{Adiabatic wave approximation}
\label{results-adwa}

\begin{table*}[t]
\begin{center}
\begin{tabular}{|c|c|c|c|c|c|c|c|c|c|c|c| r| r|}
\hline Reaction & Initial & E (MeV) & V (MeV) & r (fm) &a (fm) &W$_s$ (MeV) &r$_s$ (fm) &a$_s$ (fm) &W$_v$ (MeV) &r$_v$ (fm) &a$_v$ (fm) & $\chi^2$ & 
$\chi^2_{BG}$ \\ \hline
$^{90}$Zr(p,p) &$U_n$ &12.7 & {\it 53.99} & {\it1.249}	&{\it0.524}	&4.409	&{\it1.124}	&{\it0.790}	&1.231	&1.533	&0.573 & 0.8 & 51
  \\ \hline
$^{90}$Zr(n,n) &BG &10    & {\it50.58} &{\it1.186}	& {\it0.636}	&{\it3.334}	&1.064	&{\it0.802}	&0.600	&1.525	&0.573 & 1.7 & 83 
  \\ \hline
$^{90}$Zr(p,p) &BG & 22.5 & {\it54.79} &{\it1.139}	&{\it0.786}	&{\it6.637}	&{\it1.360}	&{\it0.659}	&2.134	&1.405	&0.590 & 1.1 & 2.7
  \\ \hline
$^{90}$Zr(d,d) & AC & 23.2 & 91.99 & {\it1.178}	&0.675	&{\it9.385}	&{\it1.284}	&{\it0.902}	&2.180	&1.042	&0.537 & 65 & 122
  \\ \hline
\end{tabular}
\caption{Best-fit optical potential parameters for elastic scattering of neutrons, protons and deuterons on $^{90}$Zr at the relevant energies, obtained with the uncorrelated $\chi^2$ function. Also shown are the $\chi^2$ values at the minimum, $\chi^2_{UC}$, and for the starting point \cite{bg69}, $\chi^2_{BG}$. Parameters in italics were allowed to vary in the fit. }
\label{tab:pot1}
\end{center}
\end{table*}

\begin{table*}[t]
\begin{center}
\begin{tabular}{|c|c|c|c|c|c|c|c|c|c|c| r| }
\hline Reaction &E (MeV) & V (MeV) & r (fm) &a (fm) &W$_s$ (MeV) &r$_s$ (fm) &a$_s$ (fm) &W$_v$ (MeV) &r$_v$ (fm) &a$_v$ (fm) & $\chi^2$ \\ \hline
$^{90}$Zr(p,p) &12.7 & {\it52.86} & {\it1.252}	&{\it0.545}	&{\it6.628} &1.282	&0.633	&0.572	&1.266	&0.709 & 0.1  
  \\ \hline
$^{90}$Zr(n,n) &10 &{\it51.89} &{\it1.160}	&{\it0.672}	&4.054	&{\it1.282}	&{\it0.633}	&0.572	&1.266	&0.709 & 0.6
  \\ \hline
$^{90}$Zr(p,p) &22.5 &{\it50.13} &{\it1.218} &{\it0.604}	&7.180	&{\it1.299}	&{\it0.694}	&1.845	&1.409	&0.610 & 0.1
  \\ \hline
$^{90}$Zr(d,d) &23.2 &90.69 &{\it1.186} &{\it0.695}	&{\it3.035}	&1.278	&0.497	&9.695	&1.150	&0.410 &0.2 
  \\ \hline
\end{tabular}
\caption{Best Fit Parameterization for correlated nucleon and deuteron elastic scattering fitting. Parameters in italics were allowed to vary in the fit.}
\label{tab:pot2}
\end{center}
\end{table*}

To illustrate this study in detail, we picked one case: $^{90}$Zr(d,p)$^{91}$Zr(g.s.) at 22.7 MeV.  In ADWA we need the optical potentials for n-$^{90}$Zr and p-$^{90}$Zr at half the deuteron energy and p-$^{90}$Zr at the energy in the exit channel (this data being more readily available than p-$^{91}$Zr). The assumption that  the nucleon optical potentials should be computed at half the deuteron beam energy has been questioned in \cite{timofeyuk2013}, however we here maintain the traditional approach.  Using the data referenced in Table \ref{tab:data} and the uncorrelated $\chi^2$ function, we obtained the best-fit parameters shown in Table \ref{tab:pot1}. For all but the $^{90}$Zr(p,p) at 12.7 MeV, we  initialized the minimization procedure with the global parameters of Becchetti and Greenlees (BG) \cite{bg69}. For the (p,p) scattering at 12.7 MeV, the BG initialization led to a minimum with a geometry very different than that obtained for the corresponding neutron optical potential. Thus, for consistency, we  used the best-fit obtained for (n,n) at 10 MeV as the starting point for the protons. In this way, the optical potentials are somewhat similar as one would expect based on physical considerations. 

The parameters obtained (shown in Table \ref{tab:pot1}) are physically reasonable, except for the radius of the imaginary volume term which is rather large. This apparent issue has no consequences because the depth of the imaginary volume term is very small.  In some cases, there were several parameters that were fixed in the final minimization procedure. Only the parameters shown in italic in Table \ref{tab:pot1}  were allowed to vary in the final state of the minimization procedure.  Generally, this is because some parameters could not be constrained by the data within this fitting method, mostly  parameters in the imaginary volume term. This is to be expected because at these energies the absorption is mostly at the surface.

The $\chi^2$ per degree of freedom for the best fit and the initial BG are shown in the last two columns of Table \ref{tab:pot1}, respectively. The quality of the fits are excellent and much better than what was obtained with the initial BG potential. We then pull from the $\chi^2$ function around the minimum to obtain the 95\% confidence bands as described in Section \ref{theory}.

This whole procedure was  repeated for the correlated $\chi^2_C$. In this case, we considered two initial starting points: the global parameterization BG \cite{bg69} and the best fit obtained for the uncorrelated case shown in Table \ref{tab:pot1}. We found that the best-fit parameters  do depend on the initialization. There are several local minima in the parameter space, and we found that the lowest $\chi^2$ is obtained for the first initialization choice. Thus, Table  \ref{tab:pot2} shows the best-fit parameters for the BG \cite{bg69} initialization.  It is worth to note that the correlations in the parameters are strongly reduced in the correlated fit when compared to the uncorrelated fit.

\begin{figure}[t!]
\begin{center}
\includegraphics[width=0.45\textwidth]{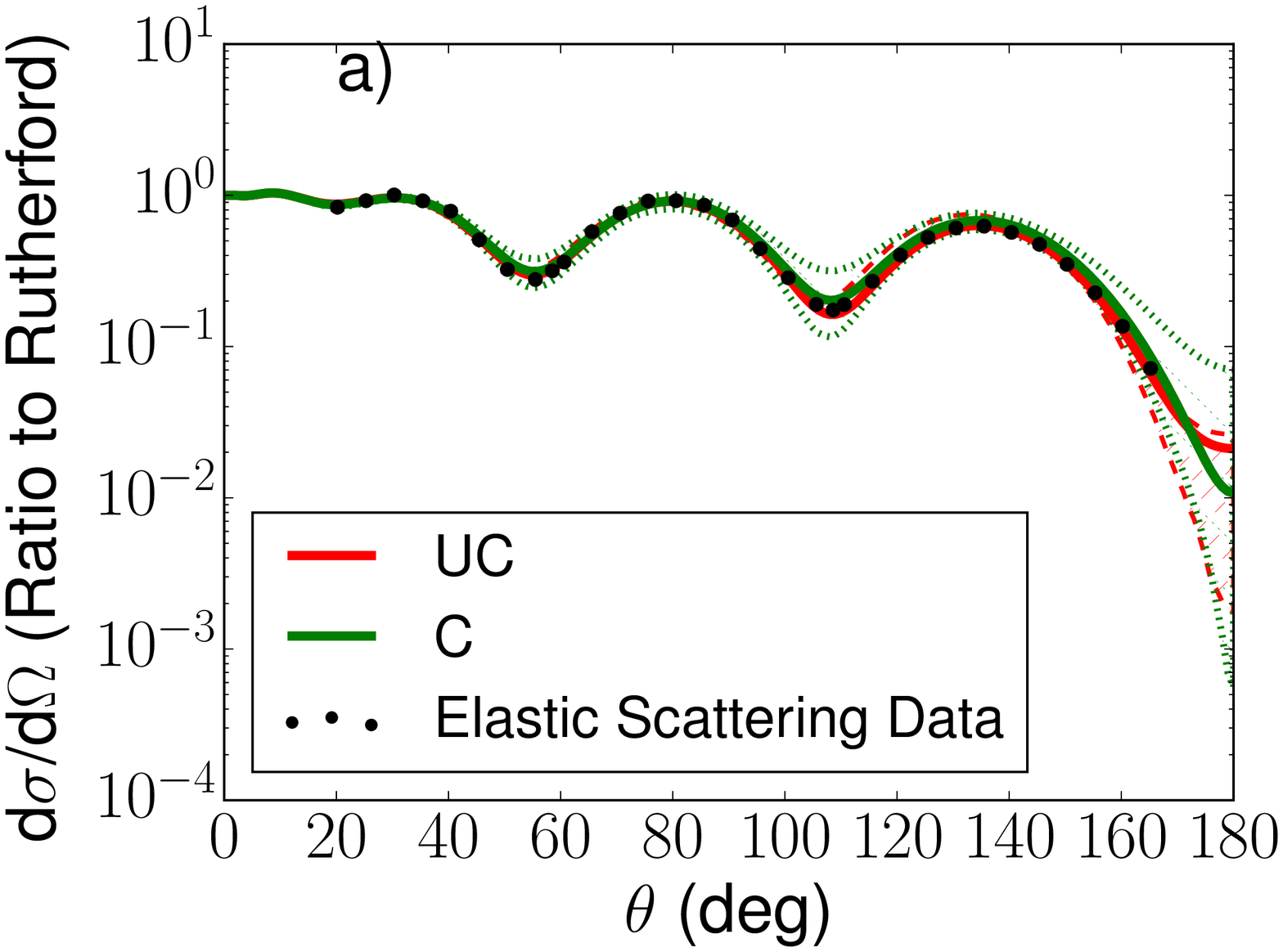}
\includegraphics[width=0.45\textwidth]{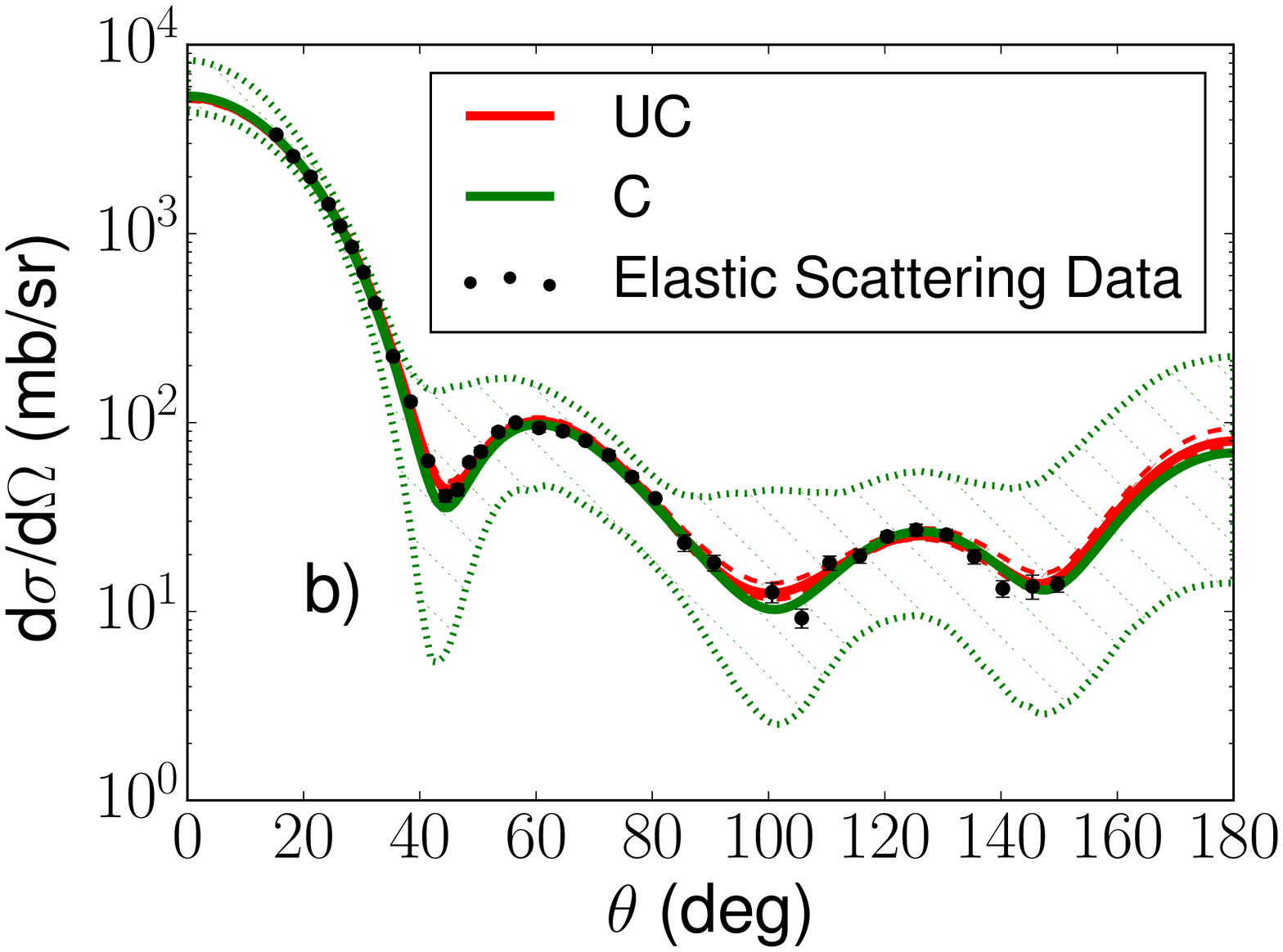}
\includegraphics[width=0.45\textwidth]{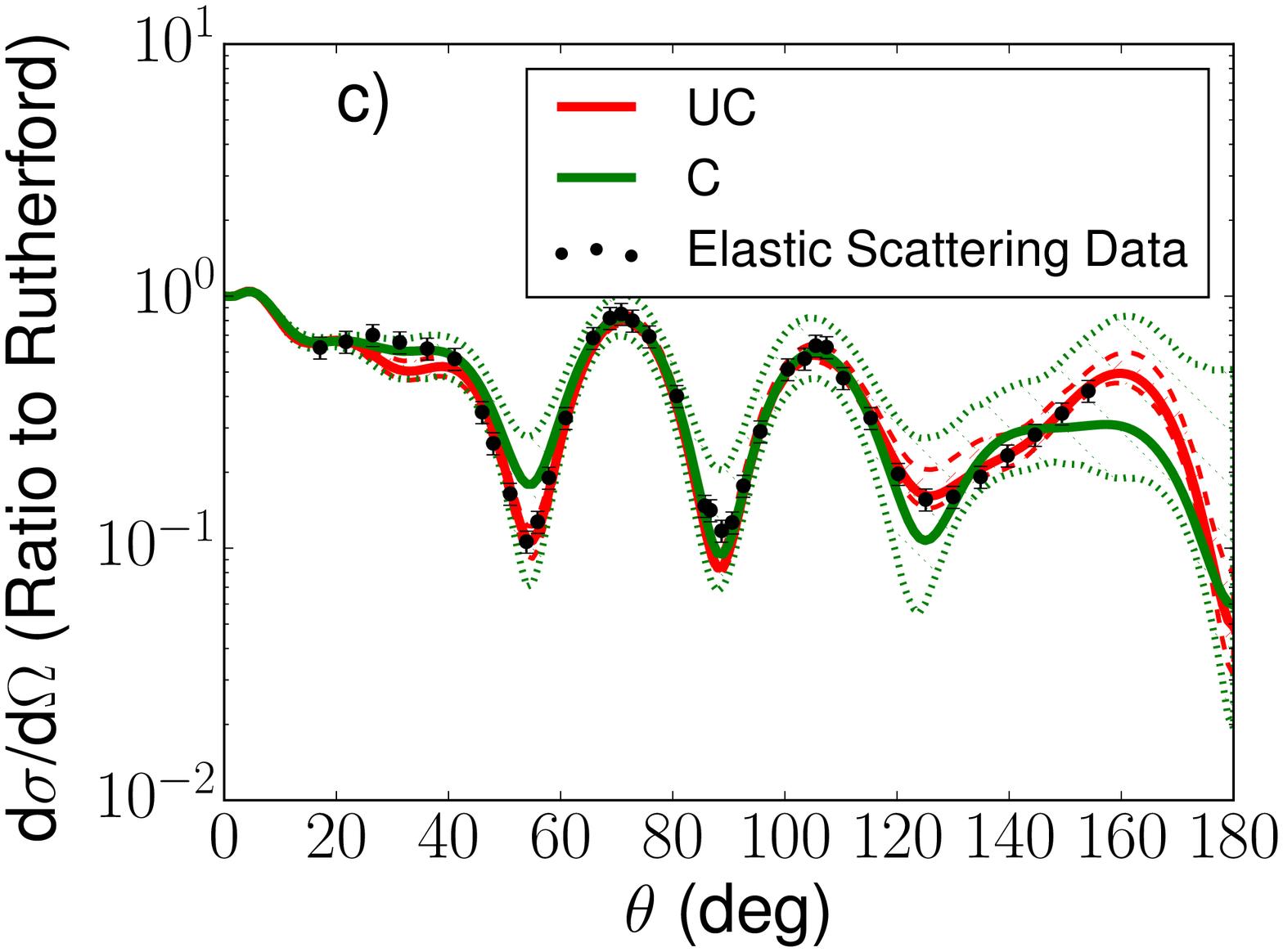}
\caption{Elastic scattering on $^{90}$Zr: (a) protons at 12.7 MeV, (b) neutrons at 10 MeV and (c) protons at 22.5 MeV; best-fit predictions using the uncorrelated $\chi^2$ (red dashed line) and correlated $\chi^2$ (green dotted line); the  95\% confidence bands using the uncorrelated (red-dashed-hashed band) and correlated $\chi^2$ (green-dotted-hashed band). Data from \cite{90zrp12,90zrn10,90zrp22}.}
\label{fig:Nel}
\end{center}
\end{figure}

The prediction of the elastic angular distributions obtained with the best fits of Tables \ref{tab:pot1} and \ref{tab:pot2} are shown in Fig. \ref{fig:Nel}:  the red dashed line for the uncorrelated case (UC), and the green dotted line for the correlated case (C). Also shown are  the corresponding 95\% confidence bands (red-dashed-hashed for the uncorrelated case and green-dotted-hashed for the correlated case) and the data. As in \cite{Lovell2017}, the bands obtained using the uncorrelated $\chi^2$ are  narrow. These become much wider when model correlations are included.

\begin{figure}[t!]
\begin{center}
\includegraphics[width=0.45\textwidth]{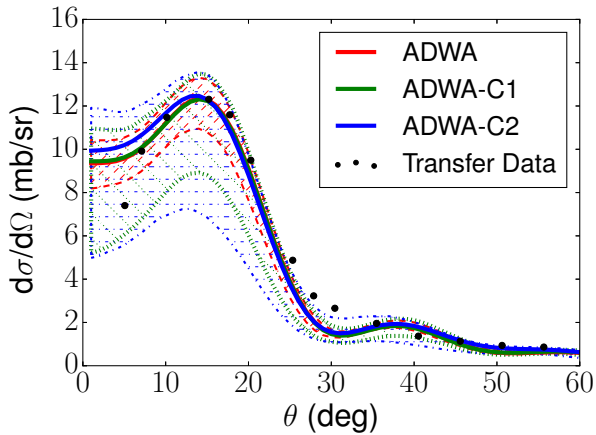}
\end{center}
\caption{ADWA angular distribution for $^{90}$Zr(d,p) at 22.7 MeV: correlated (C)  and uncorrelated (UC) 95\% confidence band predictions. Best-fit predictions are shown by the red-dashed line (uncorrelated), green-dotted line (correlated with uncorrelated initialization), blue-dot-dashed line (correlated with BG initialization). Data from \cite{90zrdpdata}.}
\label{fig:adwa}
\end{figure}

As the last step, we use these optimized optical potentials  in an ADWA calculation for the transfer cross section $^{90}$Zr(d,p) at 22.7 MeV. The ADWA angular distribution  obtained with the best fits are shown in Fig. \ref{fig:adwa} for the uncorrelated (ADWA red-dashed line) and correlated  (ADWA-C1 green-dotted line) cases. These have been normalized to the data at the peak of the angular distribution with normalization of $S=0.70$ for the uncorrelated case and $S=0.75$ for the correlated case. We see that the two ADWA angular distributions  are essentially the same - correlations in the fit do not affect the angular dependence for transfer.

Since all three nucleon potentials (neutron and proton potential in the incident channel and the proton potential in the outgoing channel) are needed to calculate the ADWA transfer cross section, we pull randomly from their corresponding $\chi^2$ distributions to obtain 95\% confidence bands for the ADWA transfer angular distributions. The results are shown in Fig. \ref{fig:adwa} with the red-dashed-hashed band corresponding to the uncorrelated case, and the green-dotted-hashed band to the correlated.  The confidence bands produced with the correlated $\chi^2$ are much wider than those obtained for the uncorrelated $\chi^2$. This could be expected given the results for elastic scattering. The band resulting from  the correlated fits is strongly asymmetric around the best-fit prediction, contrary to what is typically the case for the standard uncorrelated results.

Also added to Fig. \ref{fig:adwa} are the predictions coming from the NA best-fit parameters of the correlated $\chi^2$, initialized with the best-fit of the uncorrelated $\chi^2$  and the corresponding 95\% confidence bands (ADWA-C2 blue-dot-dashed line). The best-fit prediction provides the same angular distribution as that obtained for the uncorrelated best-fit, although the confidence band is much wider. The differences in the confidence bands ADWA-C1 and ADWA-C2 demonstrate that there is a dependence on the initialization, not only in the best-fit prediction but also on the relative width of the uncertainty predicted.

\subsection{Distorted-wave Born approximation}
\label{results-dwba}

We  studied the same reaction, $^{90}$Zr(d,p)$^{91}$Zr(g.s.) at 22.7 MeV, within DWBA. In DWBA, one needs the effective potential for d-$^{90}$Zr, in addition to the proton optical potential in the exit channel which was already studied in Section \ref{results-adwa}. Using the (d,d) elastic data at 23.2 MeV  mentioned in Table \ref{tab:data},  we obtain the best-fit parameters shown in the last row of Table \ref{tab:pot1}. We  initialized the minimization procedure with the global parameters of An and Cai (AC) \cite{ac}. The best-fit parameters obtained from the minimization are, again, physically reasonable. 
The resulting $\chi^2$ is rather large, but this is mainly due to the very small error bars on the data and  is not a good representation of the quality of the fit.
We repeated the procedure for the correlated $\chi^2$ and obtain the parameters shown in the last row of Table \ref{tab:pot2}. 

\begin{figure}[t!]
\begin{center}
\includegraphics[width=0.45\textwidth]{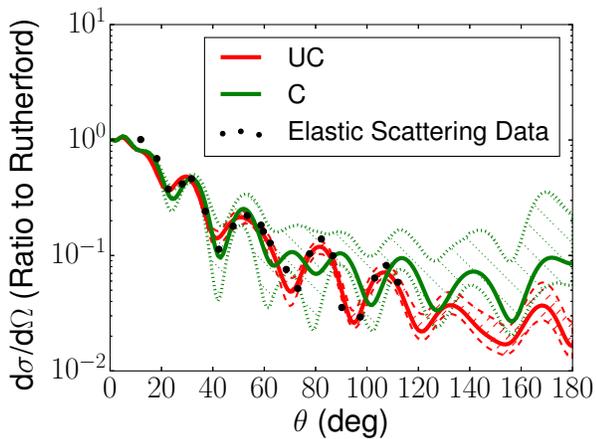}
\end{center}
\caption{Elastic scattering for $^{90}$Zr(d,d) at 23.2 MeV; best fit predictions using the uncorrelated $\chi^2$ (red-dashed line) and correlated $\chi^2$ (green-dotted line); the  corresponding 95\% confidence bands. Data from \cite{48cad23}.}
\label{fig:del}
\end{figure}

\begin{figure}[t!]
\begin{center}
\includegraphics[width=0.45\textwidth]{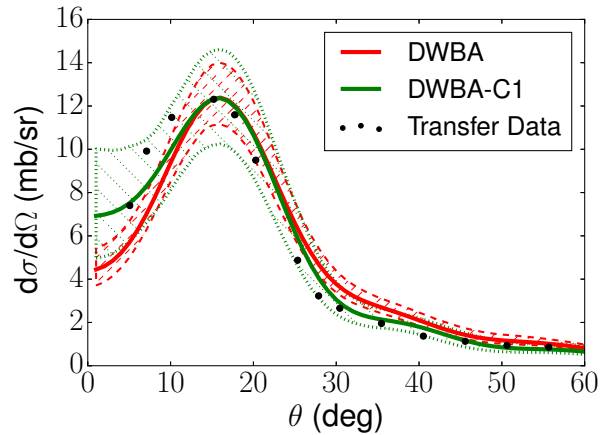}
\end{center}
\caption{Comparison of DWBA correlated and uncorrelated confidence band predictions for $^{90}$Zr(d,p) at 22.7 MeV: DWBA using the uncorrelated $\chi^2$ (red-dashed line) and DWBA-C1 using the correlated $\chi^2$ (green-dotted line). DWBA-C1 corresponds to the correlated fit  using  the global potential AC \cite{ac} as initial parameters. Data from \cite{90zrdpdata}.}
\label{fig:dwba}
\end{figure}

The elastic angular distributions for $^{90}$Zr(d,d) at 23.2 MeV, for the best-fit parameters resulting from the uncorrelated $\chi^2$ (UC red-dashed line)  and 
$\chi_C^2$ (C green-dotted line), are shown in Fig.\ref{fig:del}, and the corresponding transfer angular distribution predicted by DWBA are shown in Fig. \ref{fig:dwba} (labelled DWBA and DWBA-C1 respectively). The best-fit transfer prediction use the best fit for the deuteron optical potential in the entrance channel and the proton optical potential for the exit channel provided in Tables \ref{tab:pot1} and \ref{tab:pot2}. As in Fig.\ref{fig:adwa}, the transfer angular distributions have been normalized to the transfer data at the peak of the distribution. The normalizations are $S=0.73$ and $S=0.51$ for predictions with best-fit parameters obtained from the uncorrelated and correlated $\chi^2$ functions respectively. Note that, particularly at small angles, the best-fit angular distributions predicted with DWBA for the uncorrelated and the correlated case are considerably different.

Also shown in Fig.\ref{fig:dwba} are the corresponding 95\% confidence bands (red-dashed for the uncorrelated  and green-dotted for the correlated cases).
As for ADWA, the  band obtained from the uncorrelated $\chi^2$-functions is much narrower than the one obtained with the $\chi_C^2$-functions. 

\subsection{Comparing reaction models}
\label{compare}

\begin{table}[t!]
\begin{center}
\begin{tabular}{|c|c|c|c|c|c|c|c|}
\hline Target & Model & $\epsilon_{p,in}$  & $\epsilon_{n,in}$  & $\epsilon_{d,in}$ & $\epsilon_{p,out}$  &  $\epsilon_{dp}$\\ \hline
$^{90}$Zr & ADWA &8.1 &8.1 &-- &11 &19\\ 
$^{90}$Zr & DWBA &-- &-- &5 &11 &23\\ 
$^{90}$Zr & ADWA-C1 &75 &42 &-- &37 & 52\\ 
$^{90}$Zr & DWBA-C1 &-- &-- &17 &37 & 35\\ \hline
\end{tabular}
\caption{Relative widths of the 95\% confidence bands, $\epsilon$, at the first peak of the  angular distribution for reactions on $^{90}$Zr.}
\label{tab:error}
\end{center}
\end{table}

Next we compare directly the results obtained within DWBA and ADWA. We first consider the magnitude of uncertainty obtained within the two models.
We define the relative width of the band at a given angle $\theta$ as $\epsilon = \frac{\sigma_{max}(\theta)-\sigma_{min}(\theta)}{\sigma_{best-fit}} \times 100$.
The actual total uncertainties obtained in the transfer predictions of Figs. \ref{fig:adwa} and \ref{fig:dwba} at the peak of the distributions are given in the last column of Table \ref{tab:error}.  The first two rows of Table \ref{tab:error} correspond to results obtained with the uncorrelated $\chi^2$ and the remaining rows use $\chi_C^2$. We immediately see in the last column that, for the uncorrelated case,  the uncertainty obtained within ADWA is slightly smaller than that obtained within DWBA, while the opposite is true for the correlated results.

To explore how the uncertainties associated with the various input optical potentials propagate to the transfer, we compute the relative widths of the uncertainty bands at the first peak for all of the relevant elastic-scattering reactions on $^{90}$Zr for both ADWA and DWBA. These are also shown in Table \ref{tab:error}: columns 3,4, 5 and 6  $\epsilon_{p,in}$, $\epsilon_{n,in}$, $\epsilon_{p,out}$ and  $\epsilon_{d,in}$ are the relative widths of the uncertainty bands for the incoming proton, incoming neutron, outgoing proton and incoming deuteron elastic scattering, respectively.  
We also include the uncertainty obtained in the transfer channel, $\epsilon_{dp}$ listed in the last column of Table \ref{tab:error}, corresponding to the results shown in Figs.\ref{fig:adwa} and \ref{fig:dwba}.  

\begin{figure}[t!]
\begin{center}
\includegraphics[width=0.45\textwidth]{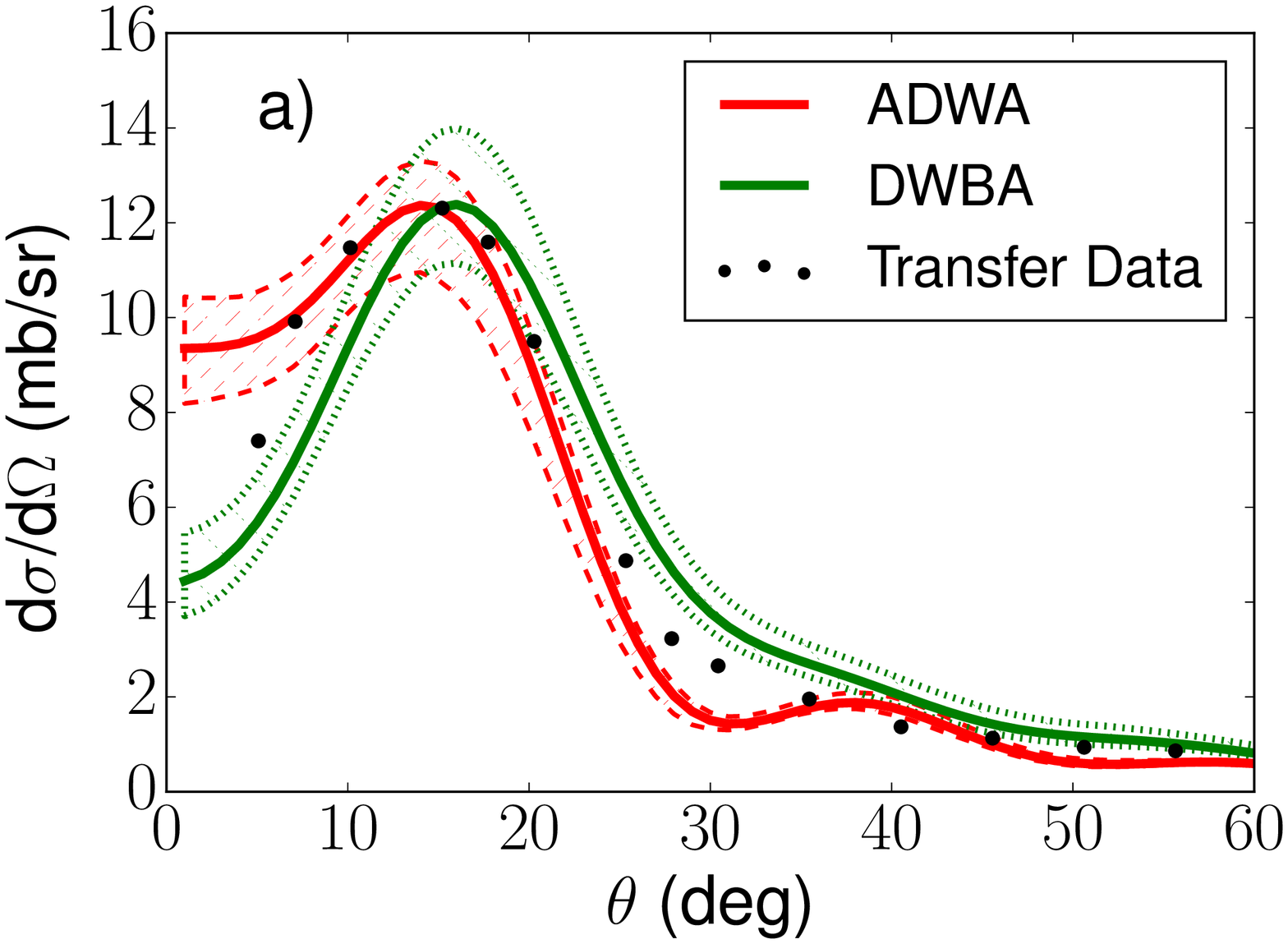}
\includegraphics[width=0.45\textwidth]{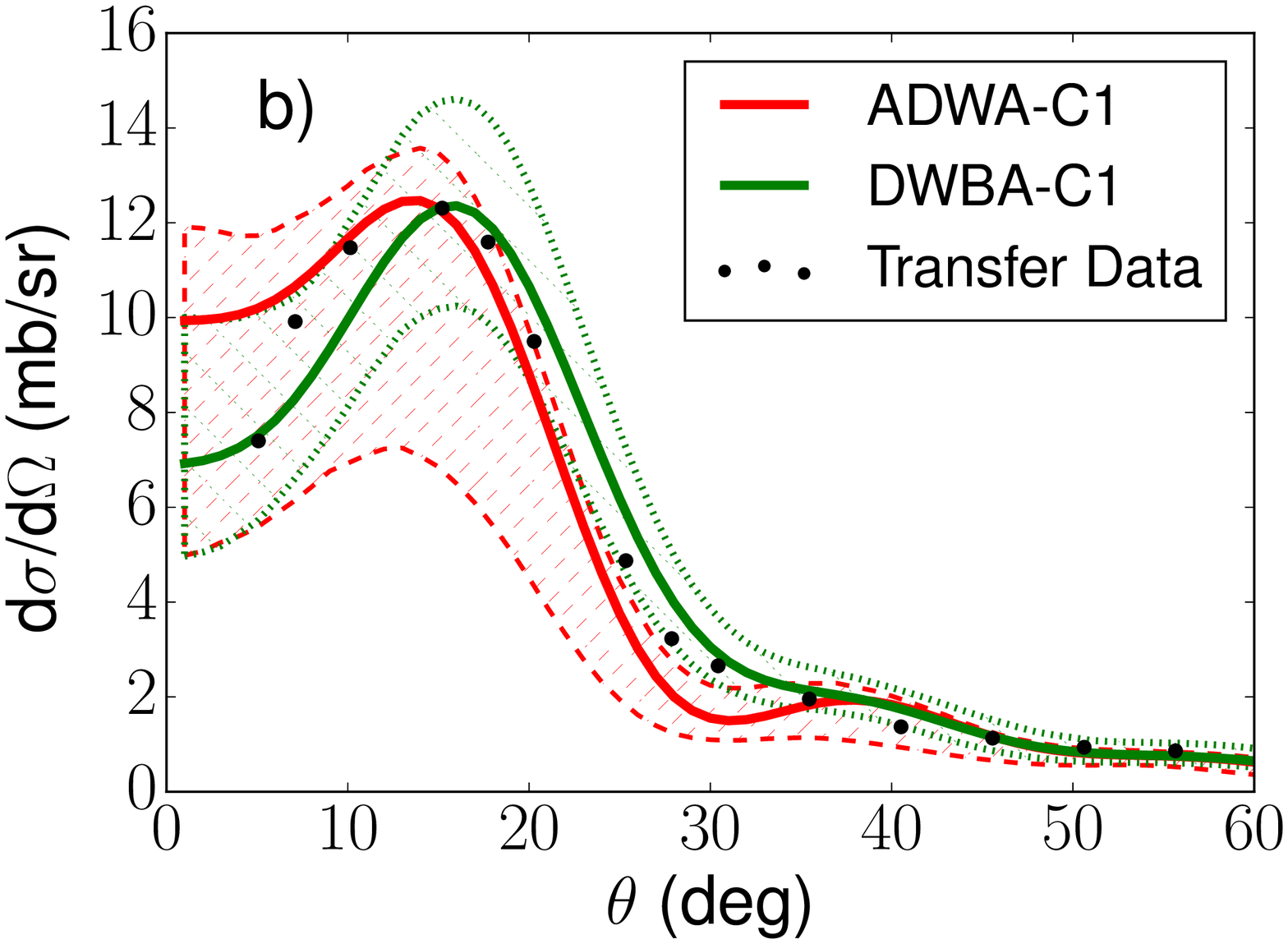}
\end{center}
\caption{Comparison of ADWA and DWBA confidence band predictions for $^{90}$Zr(d,p) at 22.7 MeV: (a) uncorrelated and (b) correlated. Data from \cite{90zrdpdata}.}
\label{fig:addw}
\end{figure}

We now compare directly the ADWA and DWBA distributions in Fig.\ref{fig:addw}, assuming no correlations (panel a) and including correlations in the minimization procedures (panel b). From the results ignoring correlations, one might be tempted to favor DWBA over ADWA, given the comparison of the angular distributions at small angles with data.
However, when including correlations, the large uncertainties quantified in Table \ref{tab:error} blur the picture. Then, both ADWA and DWBA are consistent with the data.  

\subsection{Other cases studied}
\label{other}

\begin{figure}[t!]
\begin{center}
\includegraphics[width=0.45\textwidth]{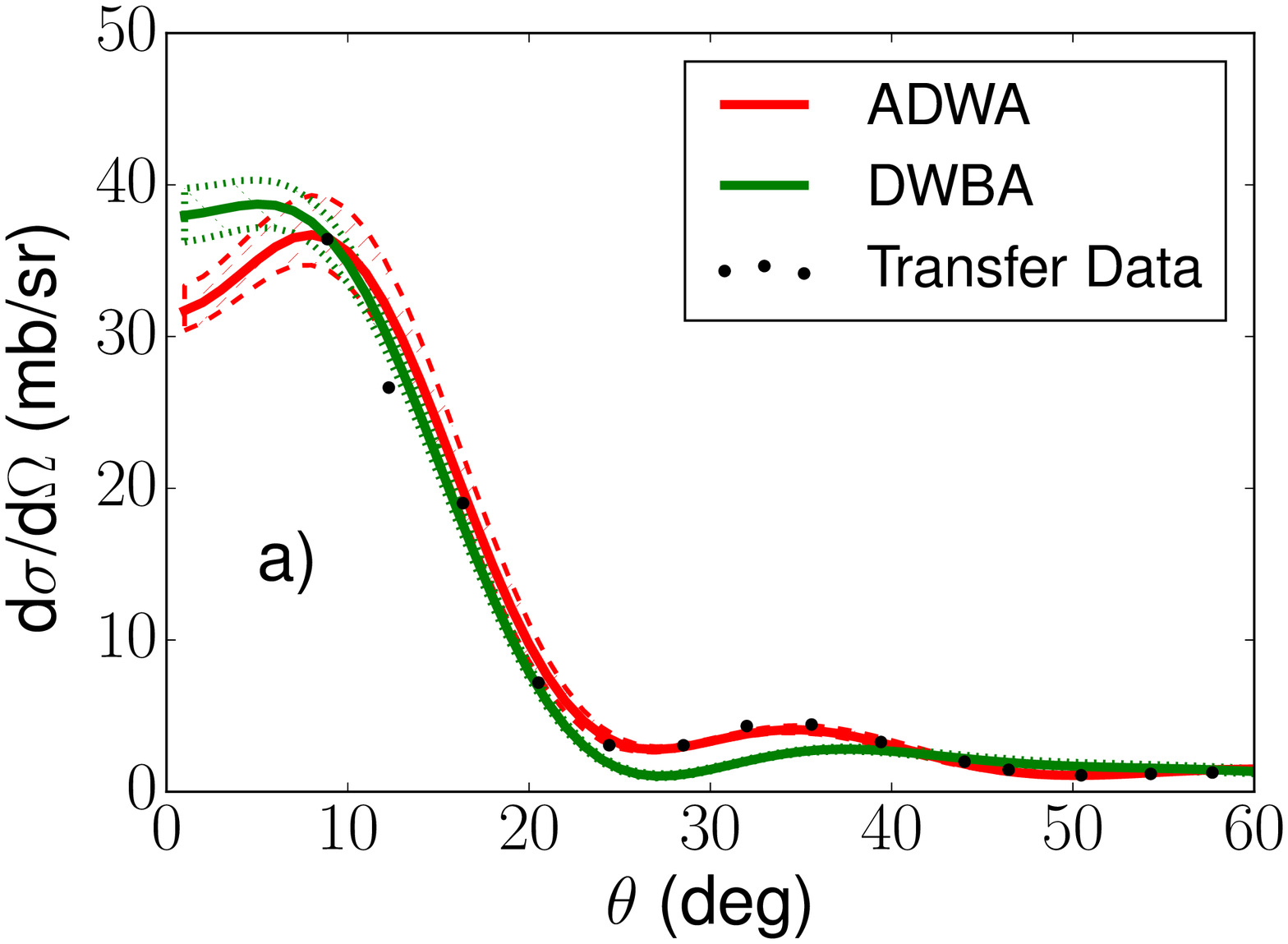}
\includegraphics[width=0.45\textwidth]{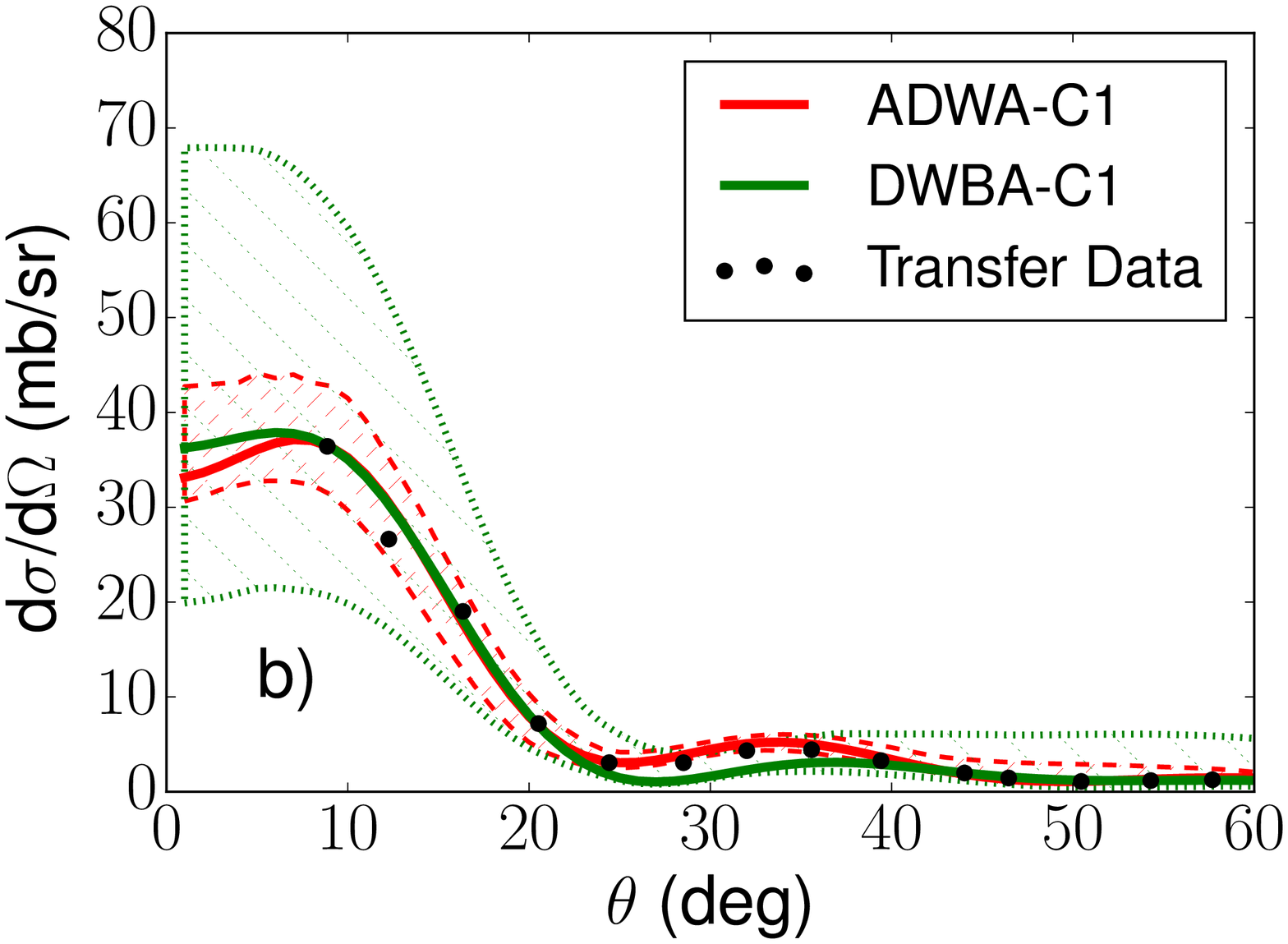}
\end{center}
\caption{Comparison of ADWA and DWBA confidence band predictions for $^{48}$Ca(d,p) at 19.3 MeV: (a) uncorrelated and (b) correlated. Data from \cite{48cadpdata}}.
\label{fig:ca}
\end{figure}
\begin{figure}[t!]
\begin{center}
\includegraphics[width=0.45\textwidth]{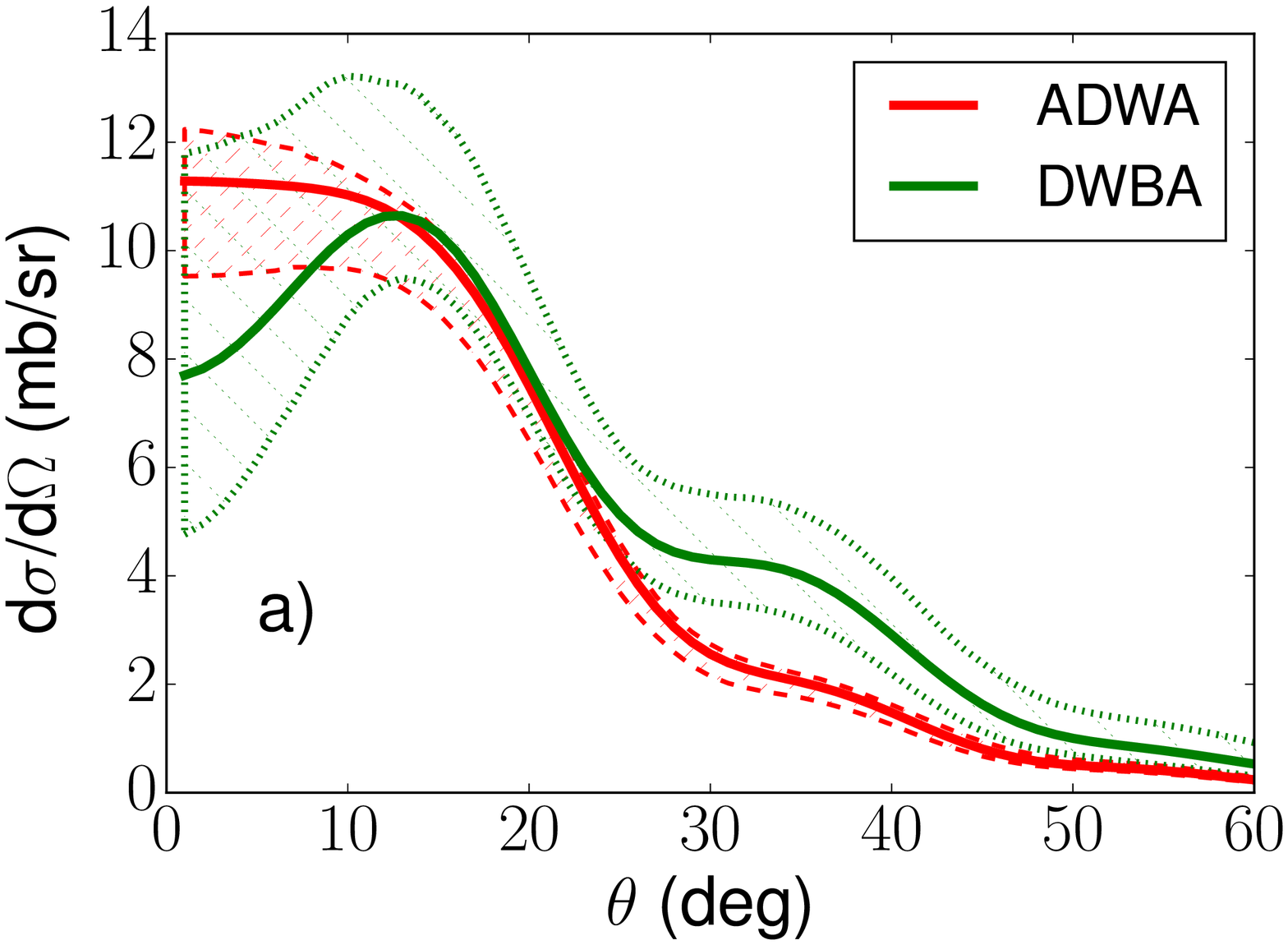}
\includegraphics[width=0.45\textwidth]{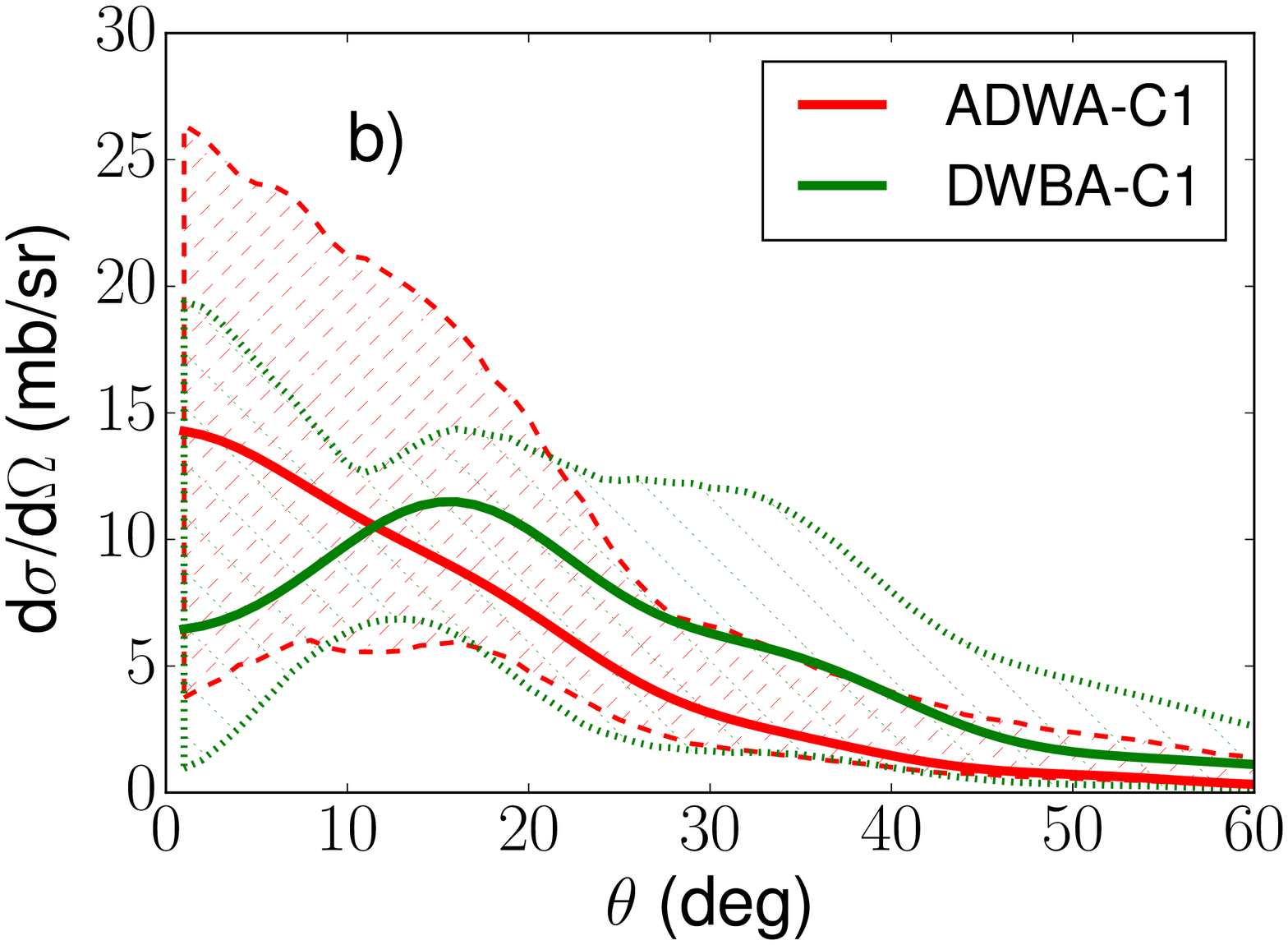}
\end{center}
\caption{Comparison of ADWA and DWBA confidence band predictions for $^{208}$Pb(d,p) at 32.9 MeV: (a) uncorrelated and (b) correlated.}
\label{fig:pb}
\end{figure}

So far we have discussed the results specifically for the $^{90}$Zr case. Since the lessons drawn can differ  when other targets are considered, here we include a summary of the results on $^{48}$Ca and $^{208}$Pb.  
Table \ref{tab:error2} contains the same relative widths as Table \ref{tab:error} for the reactions on $^{48}$Ca and $^{208}$Pb. Focusing first in the last column of Table \ref{tab:error2}, in both $^{48}$Ca(d,p) and $^{208}$Pb(d,p),  ADWA provides  a smaller uncertainty than DWBA, with the exception of uncorrelated results for $^{48}$Ca(d,p). For this case, all elastic scattering fits, for both  correlated and uncorrelated, produce narrow bands with the exception of the neutron elastic scattering and this then reflects itself in a slightly wider band for the transfer prediction in ADWA.  

 As for the $^{90}$Zr case, for both $^{48}$Ca and $^{208}$Pb, the results from the correlated fits have a wider uncertainty band than those from the uncorrelated fits. 
For $^{48}$Ca, the uncertainties estimated from quadrature are much larger than the actual uncertainties when using $\chi_C^2$. Also the deuteron elastic data on $^{48}$Ca is not as precise as for $^{90}$Zr, and therefore the deuteron channel contributes more strongly to the errors in the DWBA model, particularly for the analysis with correlations. For $^{208}$Pb, when using $\chi^2_C$, the uncertainties in the outgoing proton channel dominate the errors, and estimates of errors with quadrature provide a very large number, larger by a factor of three than the actual uncertainty obtained when performing the transfer calculations. 

\begin{table}[t!]
\begin{center}
\begin{tabular}{|c|c|c|c|c|c|c|}
\hline Target & Model & $\epsilon_{p,in}$  & $\epsilon_{n,in}$  & $\epsilon_{d,in}$ & $\epsilon_{p,out}$  &  $\epsilon_{dp}$\\ \hline
$^{48}$Ca & ADWA &10 &24 &-- &8.0 &12\\ 
$^{48}$Ca & DWBA &-- &-- &11 &8.0 &8.3\\ 
$^{48}$Ca & ADWA-C&49 &321 &-- &117  &30\\ 
$^{48}$Ca & DWBA-C&-- &-- &192 &117 &132\\ \hline
$^{208}$Pb & ADWA &2.9 & 53 &-- &23 &24\\ 
$^{208}$Pb & DWBA &-- &-- & 23 &23 & 34\\ 
$^{208}$Pb & ADWA-C &18 &100 &-- & 143 & 50 \\ 
$^{208}$Pb & DWBA-C &-- &-- &61 & 143 & 71 \\ \hline
\end{tabular}
\caption{Relative widths of the 95\% confidence bands, $\epsilon$, at the first peak of the  angular distribution for reactions on $^{90}$Zr.}
\label{tab:error2}
\end{center}
\end{table}

\begin{table}[t!]
\begin{center}
\begin{tabular}{|c|c|c|c|r|r|}
\hline target & model & E (MeV) & $\theta$ (deg)   & $\epsilon_{UC}$ (\%) & $\epsilon_{C}$ (\%)\\ \hline
$^{48}$Ca & ADWA 	& 19.3 & 8				&12  &30 \\ \hline
$^{48}$Ca& DWBA 	&19.3 & 5.0			&8.3 &132 \\ \hline
$^{90}$Zr& ADWA 	&22.7 & 14			&19 &52\\ \hline
$^{90}$Zr& DWBA 	&22.7 &16				&23 & 36  \\ \hline
$^{208}$Pb & ADWA	&32.9 & 1.0			&24 & 158 \\ \hline
$^{208}$Pb& DWBA	&32.9 & 16			&34 & 71 \\ \hline
\end{tabular}
\caption{Widths of the 95\% confidence bands for the (d,p) transfer at the peak of the angular distribution, for a variety of targets: comparing results obtained with and without correlations in $\chi^2$-function (correlated fits  started from the BG parameters).}
\label{tab:error3}
\end{center}
\end{table}

 In Table \ref{tab:error3}, we summarize the relative widths of the 95\% confidence bands obtained  for the transfer angular distributions at the peak for all cases studied, assuming that all optical potential inputs are fit to the corresponding elastic scattering data (Table \ref{tab:data}). For all cases, the uncorrelated fit provides a smaller uncertainty and in some cases the correlations introduce an order of magnitude increase in the uncertainty. 

Fig.\ref{fig:ca} and Fig. \ref{fig:pb} compare the ADWA and DWBA distributions for the corresponding (d,p) reactions ($^{48}$Ca and $^{208}$Pb, respectively). Panel (a) assumes no correlations and panel b  includes correlations in the minimization process. From these results, we can draw the same conclusions as those deduced from the analysis of the $^{90}$Zr case, namely that, due to the large uncertainties produced when we include correlations in the fitting, the data cannot discriminate between ADWA and DWBA.

\section{Conclusions}
\label{conclusions}

This study follows from the work done in \cite{Lovell2017}. Here we consider (d,p) reactions on a variety of targets and quantify the uncertainties in the predicted cross sections coming from the optical potentials. We constrain the optical potentials with the corresponding elastic scattering data, using either a standard $\chi^2$ function or a correlated version as introduced in \cite{Lovell2017}. We perform transfer (d,p) calculations within both the adiabatic wave approximation and the distorted-wave Born approximation. We discuss the results for elastic and transfer on $^{90}$Zr in detail but also present results on $^{48}$Ca and $^{208}$Pb.

We find that best-fit parameters obtained with the correlated $\chi^2$ are significantly different from those obtained   with the standard uncorrelated $\chi^2$.
Systematically, the 95\% confidence bands for elastic scattering obtained pulling from  the  correlated $\chi^2$-function  are much wider than those obtained when pulling from the uncorrelated $\chi^2$. Nevertheless, the elastic scattering results are consistent in that the bands obtained from $\chi_{UC}^2$ are contained within the bands obtained with $\chi_C^2$.

When propagating the uncertainties  to transfer reactions using $\chi_{UC}^2$, the ADWA predictions differ from the DWBA predictions at small angles. Thus, if one could ignore correlations in the model, one might discriminate between the two theories. However, once correlations are included, these differences are washed out and both DWBA and ADWA predictions corroborate the transfer data. Using the first peak of the transfer (d,p) angular distribution to extract a spectroscopic factor, as is standard in our field, we obtain consistent spectroscopic factors in ADWA and DWBA.

Although we have made assumptions as to the form of the correlations in the model, the study in \cite{Lovell2017} demonstrates that these need to be included. It would be useful to have an understanding of whether our correlated $\chi^2$ function is the best representation for the correlations in the cross section observables. Bayesian statistics may offer another path toward this goal.
 
The main conclusion from this work is that the uncertainties coming from the optical potentials, constrained by all relevant elastic scattering channels, are  too large, and it is crucial to reduce them in order to enable model comparison. Work to include a larger variety of data in the fit is in the pipeline. Then we can explore which types of data offer additional and optimal constraints.

\begin{acknowledgments}
This work was supported by the National Science
Foundation under Grant  PHY-1403906, the Stewardship Science Graduate Fellowship program under Grant No. DE-NA0002135, and the
Department of Energy under Contract No. DE-FG52-
08NA28552. This work relied on iCER and the High Performance Computing Center at Michigan State University for computational resources. 
All elastic scattering data were collected from the \href{https://www-nds.iaea.org/exfor/exfor.htm}{EXFOR} database.
\end{acknowledgments}

\bibliography{uq-adwa-f3}

\end{document}